\documentclass[pre,aps,twocolumn]{revtex4}
\usepackage{graphicx,epsf,subfigure,amsbsy,amsmath,float}
\usepackage[latin1]{inputenc}

\begin{document}
\title{Global Topology from an Embedding}

\author{Robert Gilmore,$^{1,2}$ Christophe Letellier,$^2$ and 
Nicola Romanazzi$^{1}$} 

\affiliation{$^1$Physics Department, Drexel University,
Philadelphia,  Pennsylvania 19104, USA\\
$^2$CORIA UMR 6614 - Universit\'e de Rouen, Av. de l'Universit\'e, BP 12 \\
F-76801 Saint-Etienne du Rouvray cedex, France}

\date{\today}

\begin{abstract}

An embedding of chaotic data into a suitable phase space
creates a diffeomorphism of the original attractor
with the reconstructed attractor.  Although diffeomorphic,
the original and reconstructed attractors may not be
topologically equivalent.  In a previous work
we showed how the original and reconstructed attractors can differ when 
the original is three-dimensional and of genus-one type.
In the present work we extend this result to 
three-dimensional attractors of arbitrary genus.
This result describes symmetries exhibited by the
Lorenz attractor and its reconstructions.

\end{abstract}

\pacs{PACS numbers: 05.45.+b}

\maketitle

\section{Introduction}
\label{sec:introduction}

\begin{figure}[ht]
  \begin{center}
    \includegraphics[angle=0,width=4.0cm]{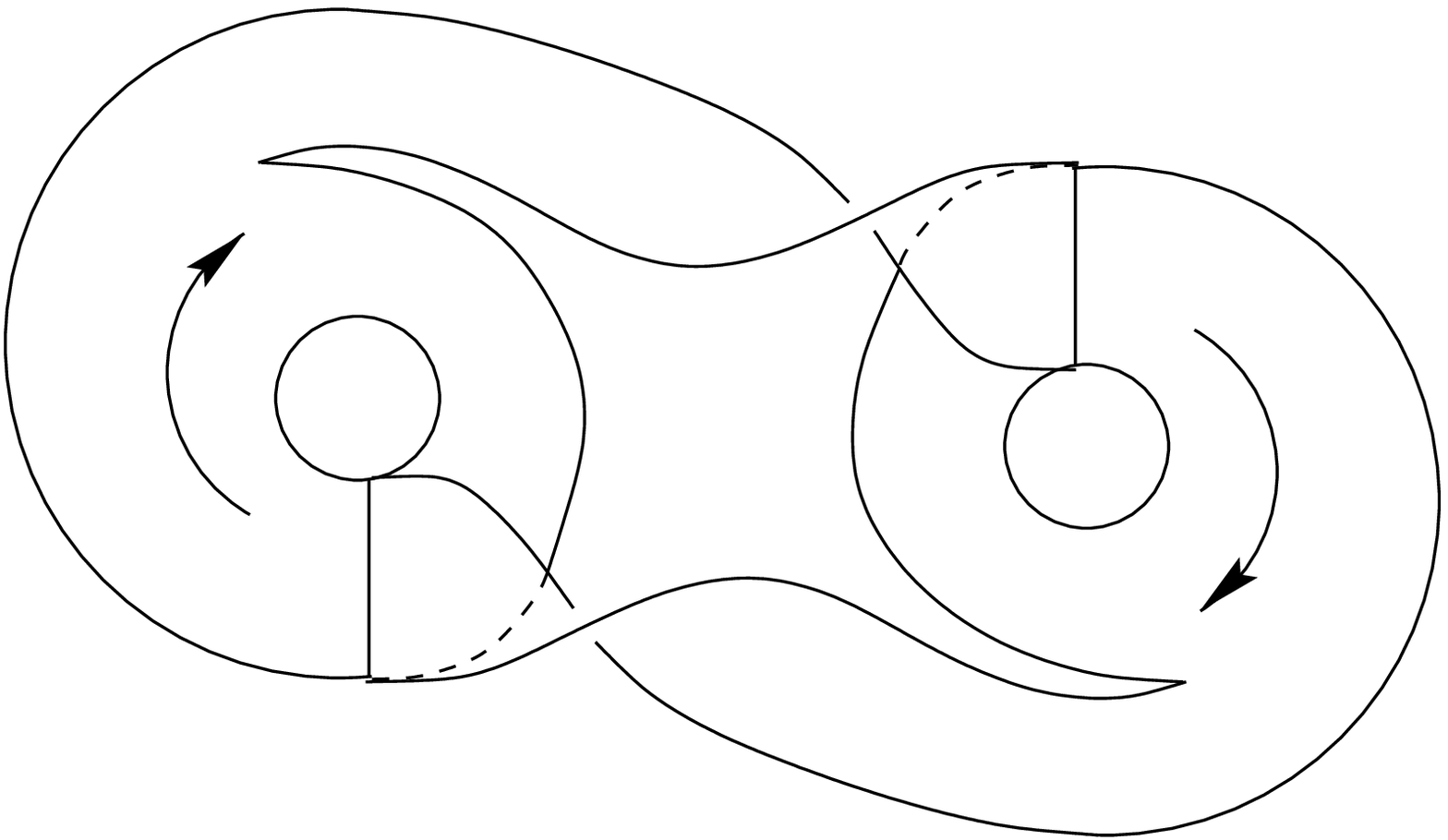} \\
    (a) With rotation symmetry \\[0.3cm]
    \includegraphics[angle=0,width=4.0cm]{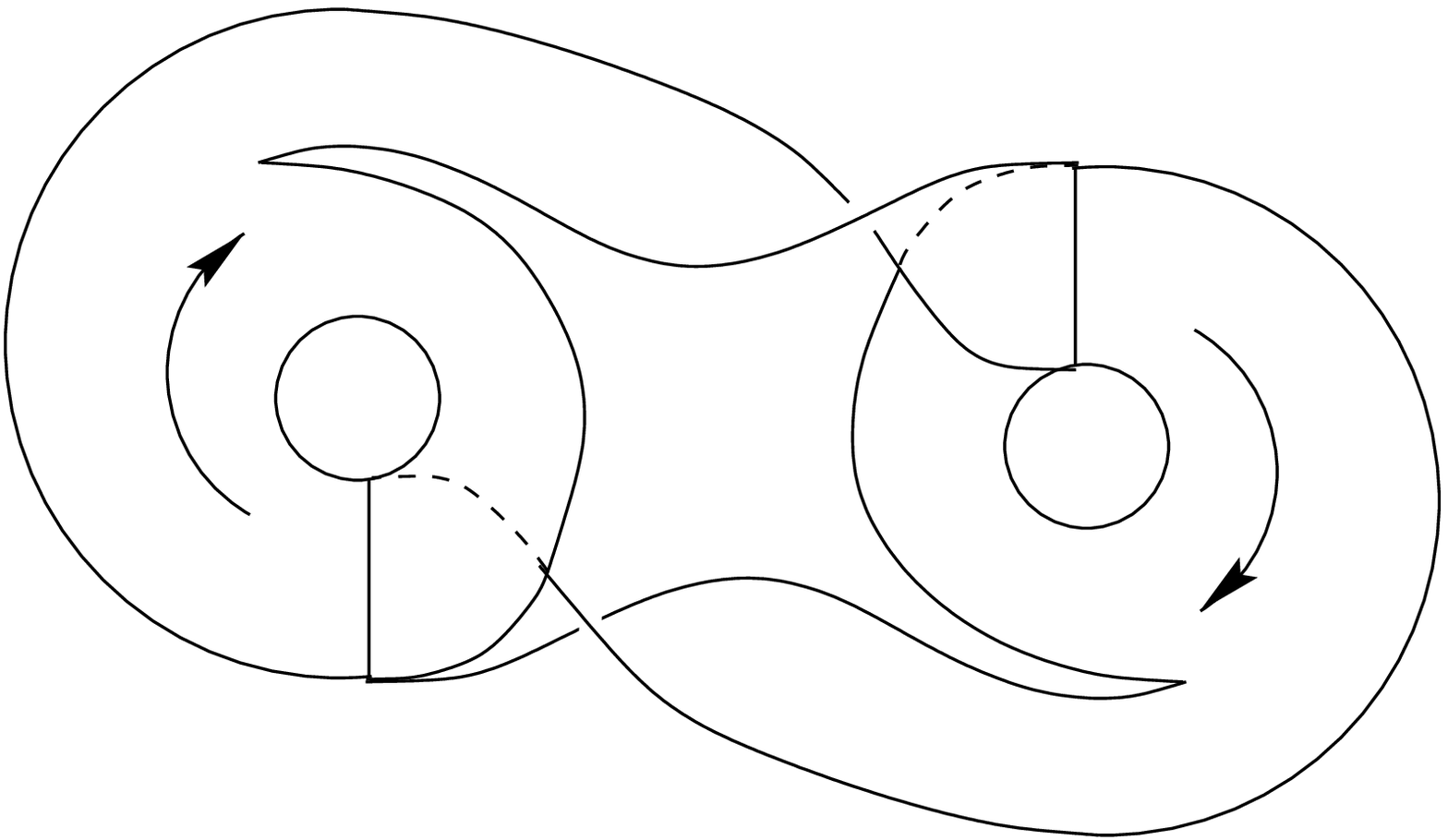} \\
    (b) With inversion symmetry \\[-0.2cm]
    \caption{Branched manifolds describing two representations of the Lorenz attractor: 
(a) with rotation symmetry; (b) with inversion symmetry.}
    \label{fig:lorenz_rep}
  \end{center}
\end{figure}

Mappings of scalar and vector time series into suitable
phase spaces are regularly used to visualize processes
that generate experimental data \cite{Pac80,Tak81}.
When the mapping is an embedding, a diffeomorphism exists 
between the original (``experimental'') attractor and the
reconstructed attractor.  It is known from numerous
examples that a single time series can be embedded
into a phase space in several different ways, giving
rise to reconstructed attractors that are diffeomorphic
but not topologically equivalent \cite{Gil02,Gil07,Tsa04b}.
By topologically equivalent (isotopic) we mean that there 
is a smooth transformation (isotopy), or change of coordinates, 
that deforms one into the other \cite{Rol90}.  If there 
is no smooth transformation that deforms one into the other, 
the two are topologically inequivalent.  As a particular example 
of topologically inequivalent attractors that are related
by a diffeomorphism restricted to the attracting set,
we cite the Lorenz attractor \cite{Lor63}.  Under the
vector embedding $(x(t),y(t),z(t)) \rightarrow R^3$ 
the attractor exhibits
rotation symmetry around the $z$-axis, but under the
scalar differential embedding $(x(t),\dot{x}(t),\ddot{x}(t))
\rightarrow R^3$ the reconstructed attractor exhibits inversion
symmetry through the origin \cite{Gil02, Gil07, Let96a, Let96b}.  
The attractors
with two different order-2 symmetries cannot be smoothly
deformed into each other in $R^3$.   Representations of these two
attractors by their branched manifolds are shown in Fig.
\ref{fig:lorenz_rep}.  It is also known that diffeomorphic
but topologically inequivalent embeddings can result from 
time delay embeddings with different delays \cite{Tsa04b}.

This raises an important question.  How much of what we learn
by studying a reconstructed attractor depends on the embedding
and how much is independent of the embedding?  The properties
that are independent of the embedding characterize the
original attractor.

Geometric properties, such as the spectrum of fractal dimensions,
are in principle diffeomorphism independent \cite{Eck85} 
(but see \cite{Lef92}).   Dynamical properties, such 
as the spectrum of Lyapunov exponents, are also 
diffeomorphism independent (but see \cite{Bro91, Sau98}).
As a result, these real numbers can usually be assumed to
be valid for the original attractor when computed from
any reconstructed attractor.  Conversely, they cannot
be used to distinguish one embedding from another.
Nor do these real numbers shed any light on the 
{\em mechanism} generating chaotic behavior \cite{Rom07}.

Topological indices shed a great deal of light on the
mechanism generating chaotic behavior 
\cite{Gil02, Gil07,Gil98}.  At the same time
they are not embedding invariants.  As a result we must
understand what part of the topological information
obtained from a reconstructed attractor is independent
of the embedding, and what part is not.  This program
has been completed for three dimensional attractors
that are contained in a bounding torus of genus one \cite{Rom07}.
In this case we find that embeddings have three degrees 
of freedom: parity, 
global torsion, and knot type.

In the present work we extend these results to three-dimensional
attractors of higher genus ($g>1$).  These include many 
attractors generated by autonomous dynamical systems with
two-fold or higher-fold symmetry \cite{Gil07,Mir93,Let01,Azi99}. 
We find the analogs of parity and global torsion, 
but do not discuss knot type, but all embeddings reveal the
same stretching and folding mechanism.

Our work is restricted to three-dimensional attractors.
These are attractors that exist in a three-dimensional 
manifold, not necessarily $R^3$.  This restriction is necessary because
the topological indices that we compute (linking numbers,
relative rotation rates) are for closed
periodic orbits that have a rigid organization in 
three-dimensional manifolds but not in higher dimensional
manifolds \cite{Rol90,Gil98}.

In Sec. II we briefly review the results for the genus-one
case.  In Sec. III we construct the analog, in the higher-genus case,
for global torsion in the genus-one case.  In Sec. IV we construct
the analog, in the higher genus-case, of parity in the genus-one
case.  We discuss the implications of our results in Sec. V.

\section{Review of Genus-One Results}
\label{sec:II}

In \cite{Rom07} we assumed that an experimental attractor is
contained in a three-dimensional manifold that has the global
topology of a genus-one torus.  An embedding constructs a 
diffeomorphism between the original and reconstructed
attractors.  A different embedding provides another
diffeomorphism between the original and another reconstructed
attractor.  The two (in fact, all) reconstructed attractors 
are diffeomorphic when restricted to the attracting
set.  The question of how embeddings of
an unseen attractor can differ simplifies to the question
of how diffeomorphisms of a torus to a torus can differ.

Diffeomorphisms form a group.  The subset of diffeomorphisms
that is isotopic to the identity forms an invariant subgroup
\cite{Rol90,Rom07}.
In fact, this invariant subgroup cannot change any 
topological indices, which are integers or rational fractions
\cite{Gil02,Gil98}.  The quotient group, 
diffeomorphisms/(diffeomorphisms isotopic to identity), is 
discrete and describes the equivalence classes of diffeomorphisms 
of the torus \cite{Rol90,Rom07}.  Each element in this discrete
group changes the topological indices in a different way.

The action of this discrete group can be understood by its action
on the boundary of the torus \cite{Rol90,Rom07}.  
This is done as follows.  Cut the torus open and stretch 
it out along the central axis.  Label the position
along the axis by an angle $\phi$, $0 \le \phi \le 2\pi$.
Choose a plane at $\phi$ and rotate the intersection of the
torus boundary with this plane by an angle $\theta$.  Set
$\theta(\phi=0) = 0$.  Now close the torus back up.  A diffeomorphism
is created by this process only when periodic boundary
conditions are satisfied, so that $\theta(\phi=2\pi) = 2 \pi n$, 
with $n$ an integer \cite{Gil07a}.  This integer is the degree of freedom
called global torsion \cite{Gil02,Rom07,Sol88}.

A parity transformation is obtained by reflecting the torus
in an external mirror.  Parity is a single index: $P= \pm 1$.

A genus-one torus can be embedded into $R^3$ by allowing its central
axis to follow the curve of any knot. We do not yet know how
to classify knots algebraically.  Even less is known about extrinsic
embeddings of higher genus tori in $R^3$. We do not discuss extrinsic
embeddings of genus-$g$ tori ($g>1$) into $R^3$ in the present work.

\section{Analog of Global Torsion}
\label{sec:III}

\begin{figure}[ht]
  \begin{center}
    \includegraphics[angle=0,width=6.0cm]{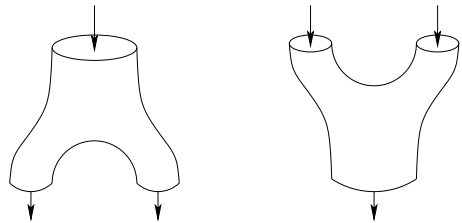} \\[-0.2cm]
    \caption{Bounding tori can be constructed from two types of units with three ports. 
(a) Splitting units have one input port and two output ports; (b) Joining units have
two input ports and one output port.}
    \label{fig:pants}
  \end{center}
\end{figure}

\begin{figure}[ht]
  \begin{center}
    \includegraphics[angle=0,width=6.0cm]{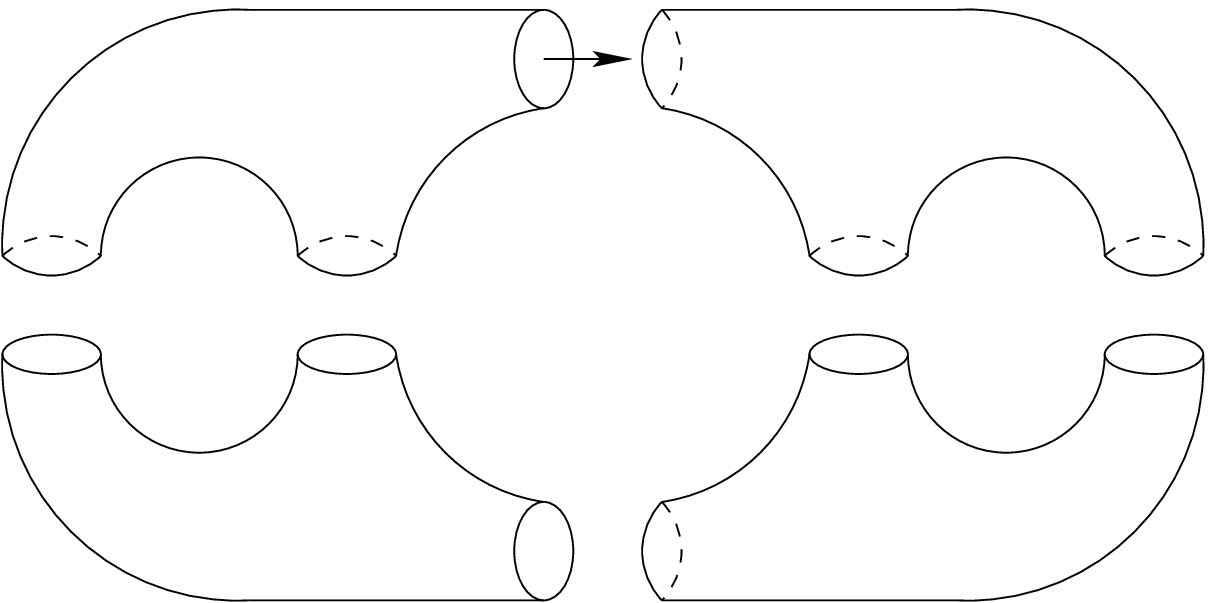} \\
    (a) \\[0.3cm]
    \includegraphics[angle=0,width=8.0cm]{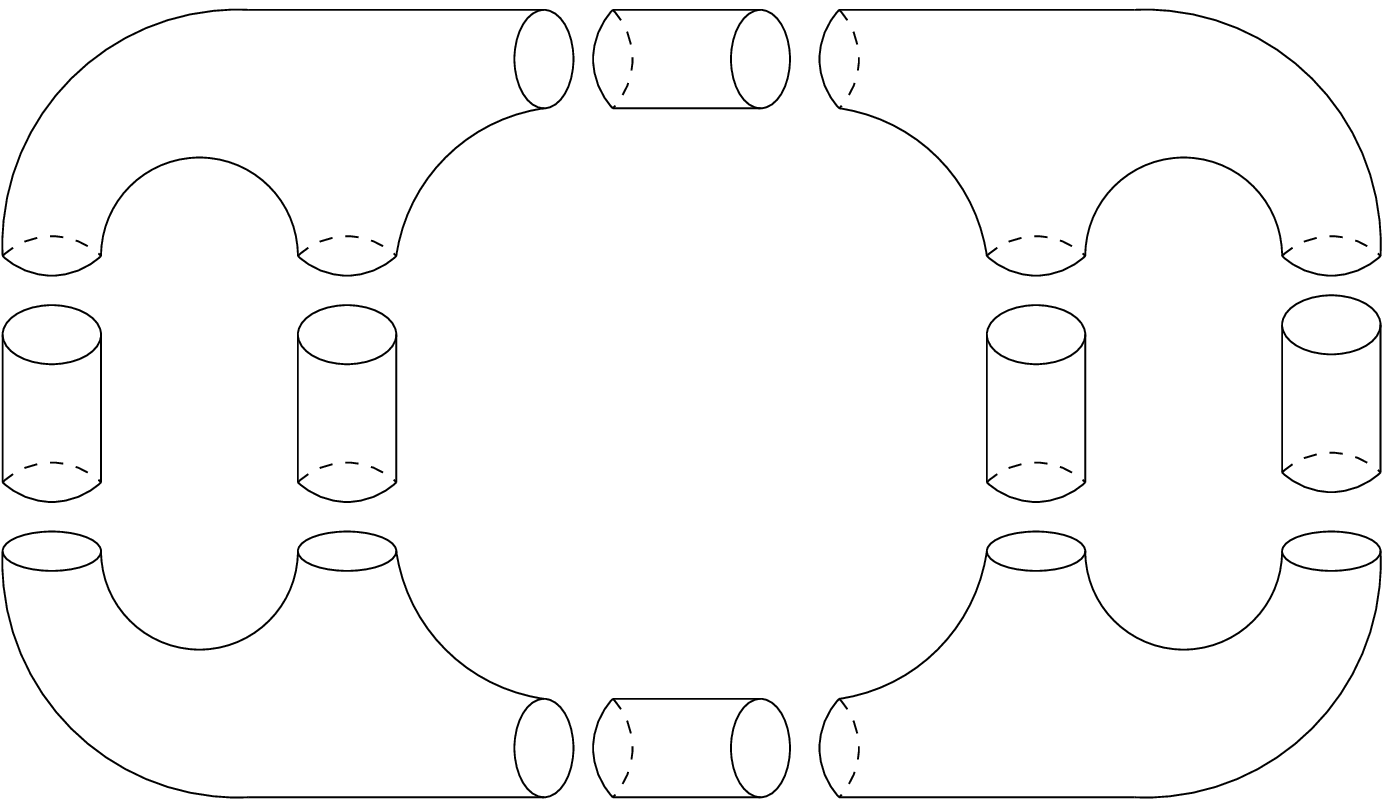} \\
    (b) \\[-0.2cm]
    \caption{(a) A genus-3 bounding torus is decomposed into
two splitting units and two joining units.
(b) Each input/output port pair is separated by a cylindrical
flow tube.  Periodic boundary conditions are satisfied if each 
flow tube is given an integer twist.  There are 6 flow tubes, 
so the analog of global torsion in the genus-3 case is $Z^6$.
In the genus-$g$ case the analog is $Z^{3(g-1)}$.}
    \label{fig:tubes}
  \end{center}
\end{figure}

A bounding torus of genus $g$ \cite{Tsa03,Tsa04a}
can be constructed, Lego$^{\copyright}$ fashion,
from $Y$-junctions.  These are two-dimensional manifolds
with three ports.  For our purposes there are two types:
splitting units with one input port and two output
ports and joining units with two input ports and one
output port.  These units are shown in Fig. \ref{fig:pants}(a)
and Fig. \ref{fig:pants}(b).  A canonical bounding torus
of genus three is shown in Fig. \ref{fig:tubes}.  The Lorenz
attractor is contained in a bounding torus of this type.
The figure shows how this bounding torus is decomposed into
two input units and two output units.  As usual, output
ports connect to input ports, and there are no free ends
\cite{Gil02, Gil98,  Bir83a, Bir83b}.

In Fig. \ref{fig:tubes}(b) we insert a ``flow tube'' between each output port
and the input port on a different unit that it is connected to.
Periodic boundary conditions are satisfied if each of these tubes is
rotated through an integer number of full twists \cite{Rom07, Gil07a}.  
Since there
are $4 = 2(3-1)$  units in the decomposition of the genus-three
torus, each has three ports, and one tube is inserted between
each pair of ports, there is a total of $(3-1) \times 3$ tubes, each of
which can exhibit an integer twist.  Each configuration is
diffeomorphic  but not isotopic to every other.

The general result is that a genus-$g$ torus can be decomposed
into $g-1$ splitting units and $g-1$ joining units, so that
$2(g-1)\times(3/2)=3(g-1)$ tubes can be inserted.  As a result, the genus-$g$
analog of the genus-$1$ global torsion is an index $Z^{N}$, $N=3(g-1)$.
This is a set of $N=3(g-1)$ integers, one for each inserted flow tube.
Recall that for bounding tori, $g=1$ or $g \ge 3$ \cite{Tsa03,Tsa04a}.

\section{Local Reflections}
\label{sec:IV}

The genus-$g$ analog of the parity transformation in the
genus-1 case consists of local reflections.

The construction of local reflections is subtle.
It is clear what a local reflection does to a
branched manifold that describes a genus-$g$ flow.
It simply maps a joining unit of a branched
manifold into its mirror image.
This is illustrated in Fig. \ref{fig:lorenz_rep}.
The problem is that local reflections in $R^3$ cannot
be used in general to create diffeomorphisms
between the two flows responsible for the branched
manifolds related by a local reflection, as shown
in Fig. \ref{fig:lorenz_rep}.  The exception occurs
when a symmetry is involved \cite{Gil07}.

\begin{figure}[ht]
  \begin{center}
    \includegraphics[angle=0,width=8.5cm]{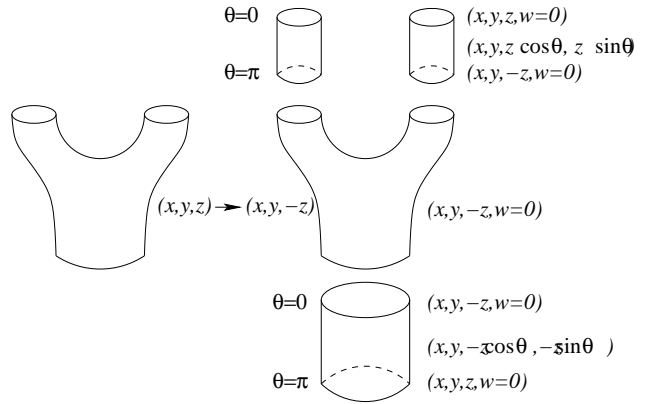} \\[-0.2cm]
    \caption{Three flow tubes are attached to a joining unit.
The flow in the joining unit undergoing local reflection
is immersed in $R^4$ according to
$(x,y,z) \rightarrow (x,y,-z,w=0)$.  The flow in the
flow tubes is rotated in the $(z,w)$ plane according to
$(z,w) \rightarrow (z \cos \theta -w \sin \theta,
z \sin \theta + w \cos \theta)$, where $\theta =0$
at the entrance of each added flow tube and $\theta = \pi$
at the exit.   This creates a diffeomorphism between
the original flow in $R^3 \subset R^4$ and a nonisotopic flow
in a three-dimensional manifold $M^3 \subset R^4$.}
    \label{fig:4d_rotation}
  \end{center}
\end{figure}

We can create diffeomorphisms that include local reflections
as shown in Fig. \ref{fig:4d_rotation}.  
Choose a joining unit and insert a
flow tube of length $L$ at each port.  Each flow tube
contains a branch of the branched mainfold describing 
the attractor generated by the flow. Deform the flow 
so that it is ``laminar'' or ``uniform'' in each flow tube.
By ``laminar'' or ``uniform'' we mean the flow assumes
the form $\dot{x}={\rm const},~ \dot{y}=0, ~\dot{z}=0 $
in local coordinates.  Here $x$ is a coordinate along 
the central axis of the cylindrical flow tube, $y$
is a coordinate in the plane of the branch through
the flow tube, and $z$ measures distance above or
below this plane.  The branch occurs in the plane
$z=0$.

Now embed the three dimensional flow into $R^4$
by introducing a fourth coordinate, $w$.
The original three dimensional flow has coordinates
$(x(t),y(t),z(t),w)$ with $w=0$.  
Now create a diffeomorphism between this flow in 
a three-dimensional manifold in $R^4$, $R^3 \subset R^4$,
and another three-dimensional manifold in $R^4$, $M^3 \subset R^4$,
as follows.  Perform a rotation through $\pi$ radians
in the $z,w$ plane in each flow tube according to 
$(z,0) \rightarrow (z \cos(x/L),z\sin(x/L))$.
This rotation maps coordinate $(y,z)$ at the
input side of a flow tube ($x=0$) to coordinate $(y,-z)$
at the output side ($x=L$).  In the joining unit, map coordinates
$(x,y,z)$ to their mirror images $(x,y,-z)$ in the
$z=0$ plane.  This set of transformations creates
a diffeomorphism between flows in $R^3$ and $M^3$.  
The projection of the branched manifold describing the
flow in $M^3$ into $R^3$ differs from the branched manifold
describing the flow in $R^3$ by the mirror image of the
joining unit, as shown in Fig. \ref{fig:lorenz_rep}.
The two branched manifolds are 1-1, locally isomorphic, and
not isotopic (i.e., globally distinct).
The flows in $R^3$ and $M^3$ are diffeomorphic but
the projection of the flow in $M^3 \subset R^4$ into $R^3$ 
is not an embedding.  This phenomenon has already been
encountered in descriptions of autonomous coupled
dynamo systems \cite{Mor07}.

Local reflections can be carried out independently on each of the
$g-1$ joining units.  The effect of a local reflection
can be seen by comparing the two representations of the
Lorenz flow shown in Fig. \ref{fig:lorenz_rep}.  A local
reflection has been carried out on a joining unit
in Fig. \ref{fig:lorenz_rep}(b).  This operation transforms
a rotation-symmetric representation of the attractor
(Fig. \ref{fig:lorenz_rep}(a)) to
an inversion-symmetric representation of the attractor
(Fig. \ref{fig:lorenz_rep}(b)).
We can describe the two representations shown in 
Fig. \ref{fig:lorenz_rep} as $(+,+)$ and $(-,+)$,
with the positions referring to the joining units on the
left and right, and the signs referring to a reflection
($-$) or no reflection ($+$).  Two other representations
are easily constructed with signatures $(-,-)$ and $(+,-)$.
The latter two are related to the former two by a global reflection
transformation.

A strange attractor in a genus-$g$ torus has $2^{(g-1)}$
representations related by local reflections.  They are all
related to each other by diffeomorphisms acting in $R^4$.
None is isotopic to any other.

\section{Summary}
\label{sec:V}

Embeddings based on scalar or vector time series
create diffeomorphisms between the original attractor
and the reconstructed attractor.  Different embeddings
create diffeomorphic reconstructed attractors that are
not necessarily topologically equivalent - that is,
not isotopic.  Since topology indicates clearly what are
the mechanisms (stretching, folding, tearing, squeezing)
that generate complex behavior \cite{Gil98}, it is an
important question to ask:
How much do we learn about the original attractor
by carrying out a topological analysis of a reconstructed
attractor, and how much about the embedding do we learn?
For the genus-one case the result is that embeddings
can differ by three degrees of freedom: parity,
global torsion, and knot type.  The mechanism displayed
is independent of the embedding \cite{Rom07}.

In this work we have answered this question for attractors
contained in higher-genus bounding tori.  We have done this by
constructing a discrete classification of all nonisotopic
(topologically inequivalent) diffeomorphisms of a 
bounding torus into itself.  We have enumerated the degrees
of freedom, not including how the bounding torus can be embedded
into $R^3$.  There are two degrees of freedom: local torsion
in each of $3(g-1)$ flow tubes and local reflections in
each of $g-1$ joining units.  It is useful to regard these
degrees of freedom as follows:  There are $2^{(g-1)}$
topologically inequivalent representations of an attractor
related to each other by different subsets of local
reflection transformations.  Each is the patriarch
for a $3(g-1)$ parameter family of strange attractors
defined by an index $Z^{3(g-1)}$ \cite{Gil07a}. 
All representations are topologically inequivalent.

What is an invariant of an embedding, and the same
for each of the $2^{(g-1)} \otimes Z^{3(g-1)}$ 
representatives of a strange attractor is the mechanism that
generates the dynamics.  The mechanism describes how the flow
is split apart to flow to different regions of the phase space,
and how different parts of the phase space are joined
\cite{Gil02,Rom07}.
This information is encoded in the transition matrix:
stretching is described by the rows of this matrix and
squeezing by the columns of this matrix \cite{Tsa03,Tsa04a}.

\acknowledgements

R. G. thanks CNRS for the invited position at CORIA for 2006-2007.

\end{document}